\documentclass[11pt,letterpaper]{article}
\pdfoutput=1
\usepackage{jheppub}
\usepackage{bbm}
\usepackage{mathrsfs}
\usepackage{slashed}
\usepackage{caption}
\usepackage{epstopdf}
\usepackage[normalem]{ulem}
\usepackage[bottom]{footmisc}
\usepackage{subcaption}
\usepackage{bbold}
\usepackage{titlesec}
\usepackage{threeparttable}
\usepackage{booktabs}
\usepackage{changepage}
\usepackage[utf8]{inputenc}

\usepackage{grffile}

\usepackage{graphicx}  
\usepackage{dcolumn}   
\usepackage{bm}        
\usepackage{amssymb}   
\usepackage{setspace}
\usepackage{amsmath, amssymb, setspace}
\usepackage{array}
\usepackage{booktabs}
\usepackage{caption}
\usepackage{indentfirst}
\usepackage{float}
\usepackage{lmodern}
\usepackage{multirow}
\usepackage{soul}
\usepackage[normalem]{ulem}

\newcommand{\SU}[1]{\ensuremath{\mathrm{SU}(#1)}}
\newcommand{\U}[1]{\ensuremath{\mathrm{U}(#1)}}
\newcommand{\Z}[1]{\ensuremath{\mathbbm{Z}_{#1}}} 

\newcommand{\Hu}{\ensuremath{H_u}}
\newcommand{\Hd}{\ensuremath{H_d}}

\DeclareMathOperator{\Tr}{Tr}

\titleformat{\subsubsection}
 {\normalfont\fontsize{12}{17}\itshape}{\thesubsubsection}{1em}{}
 
\title{Baryogenesis, Dark Matter, and Flavor Structure in Non-thermal Moduli Cosmology}

\author[a]{Mu-Chun Chen}
\author[b]{,~Volodymyr Takhistov}

 \affiliation[a]{Department of Physics and Astronomy, University of California, Irvine\\
 Irvine, CA 92697-4575, USA}
  \affiliation[b]{Department of Physics and Astronomy, University of California, Los Angeles\\
 Los Angeles, CA 90095-1547, USA}

\emailAdd{muchunc@uci.edu}
\emailAdd{vtakhist@physics.ucla.edu}
 
\abstract{
The appearance of scalar/moduli fields in the early universe, as motivated by string theory, naturally leads to non-thermal ``moduli cosmology''.~Such cosmology provides a consistent framework where the generation of radiation, baryons, and dark matter  can occur while maintaining successful Big Bang Nucleosynthesis and avoiding the cosmological moduli problem. We present a relatively economical construction with moduli cosmology, building on a variety of string-inspired components (e.g.~supersymmetry, discrete symmetries, Green-Schwarz anomaly cancellation).~We address a range of outstanding problems of particle physics and cosmology simultaneously, including the fermion mass hierarchy and flavor puzzle, the smallness  of neutrino masses, baryogenesis and dark matter. Our setup, based on discrete $\mathbbm{Z}_{12}^{R}$ symmetry and anomalous $\mathrm{U}(1)_A$, is void of the usual issues plaguing the Minimal Supersymmetric Standard Model, {\it i.e.} the $\mu$-problem and the overly-rapid proton decay due to dimension-4,-5 operators. The model is compatible with $\mathrm{SU}(5)$ Grand Unification.  The smallness of Dirac neutrino  masses is automatically established by requiring the cancellation of mixed gravitational-gauge  anomalies. The decay of the moduli field provides a common origin for the baryon number and dark matter abundance, explaining the observed cosmic coincidences, $\Omega_{B} \sim \Omega_{DM}$.  
}

\begin{document}
 \maketitle
\flushbottom


\section{Introduction}

Supersymmetry (SUSY), the maximal possible extension of the Poincare symmetry,~remains one of the most appealing solutions to the gauge hierarchy problem of the Standard Model (SM). Among its salient features are the gauge coupling unification and the natural existence of dark matter (DM) candidates, the neutralinos.

It is a difficult task to construct consistent cosmological scenarios based on standard thermal production of baryons as well as dark matter, allowing one to address the ``cosmic coincidence puzzle'' why their observed abundances are similar~\cite{Aghanim:2018eyx}, i.e. $\Omega_{DM} \sim \Omega_B$, as well as the matter-antimatter asymmetry. A non-thermal cosmology with an intermediate matter-dominated epoch can be a viable setting to resolve these issues.~From string theory, which generically requires supersymmetry at some scale for consistency, one expects existence of multiple scalar fields (moduli) with flat potentials. They acquire mass through non-perturbative effects after SUSY breaking.
During inflation moduli fields develop large values and come to dominate the energy density of the universe as they coherently oscillate near the bottom of their potential. This naturally leads to  a non-thermal ``moduli cosmology''. However, if the moduli are very light, their late-time decay on-sets radiation-dominated universe at energies below the scale allowed by the Big Bang Nucleosynthesis, leading to the  ``cosmological moduli problem'' \cite{Coughlan:1983ci,Banks:1993en}.
Making the moduli massive (i.e. ``stabilizing moduli'') 
\cite{Giudice:1998xp,Randall:1998uk,Kachru:2003aw,Balasubramanian:2005zx,Conlon:2005ki}
allows to avoid this. On other hand, massive moduli decaying to gravitinos can result in ``moduli-induced cosmological gravitino problem''~\cite{Ibe:2006am,Joichi:1994ce}. 

An attractive model with non-thermal moduli cosmology that addresses baryon-DM coincidence as well as baryogenesis, based on discrete $\Z{9}$ symmetry, was proposed by Kitano, Murayama and Ratz (KMR)~\cite{Kitano:2008tk} (for earlier related works see~\cite{Dimopoulos:1987rk,Thomas:1995ze,Cline:1990bw,Fujii:2002aj,Moroi:1999zb}).
 Here, a heavy scalar $\phi$ field, which is the scalar component of a chiral supermultiplet
$\Phi = (\phi, \tilde{\phi}, F_{\phi})$,
dominates the energy density of the early universe during its coherent oscillations 
and then later decays. The resulting radiation influx from decay washes out any dangerous relics that are present, such as the gravitinos. Enforced  $R$-parity in the model prevents direct $\phi$ decays into gravitino pairs, thus avoiding the possible gravitino problem. On the other hand, 
through the operator $\Phi \overline{U} \overline{D} \overline{D}$ present in the model,  the decay of the $\phi$ field produces baryons as well as superpartners, for example through $\phi \rightarrow q q \tilde{q}$. The decay transfers any initial $\phi$-number asymmetry into baryon number, explaining the baryon asymmetry. Additionally, the DM candidate, which is a Wino or Higgsino, is also produced by the $\phi$ decays in an appropriate amount. Since these processes happen right before the BBN, the problem of overproducing the dangerous relics is avoided. Since DM and baryon asymmetry have a common origin, the ``cosmic coincidence puzzle'' is naturally addressed in this setup
(see \cite{Fujii:2002aj,Dodelson:1991iv,Laine:1998rg,Kitano:2004sv,Kuzmin:1996he} for
some alternative proposals).
While the mechanism put forth by KMR produced a compelling cosmological setting, a variety of significant problems, particularly those related to particle physics, remained unsolved. Among the outstanding issues is the $\mu$ problem of the Minimal Supersymmetric Standard Model (MSSM), the flavor puzzle and neutrino mass generation. 

In this work we aim to construct a more complete model for non-thermal moduli cosmology, while addressing several of the major issues related to particle physics and the Standard Model. In particular, we incorporate the attractive KMR mechanism to produce a common origin for baryogenesis and DM. We address the $\mu$ problem of the MSSM and the issue of fermion masses, including the neutrinos. As has been shown in~\cite{Chen:2012jg}, only $R$ symmetries can forbid the $\mu$ term. This motivates us to base our construction on an anomaly free (ensured by 
the 4-dimensional Green-Schwarz mechanism~\cite{Green:1984sg}) discrete
$R$-symmetry. Further, we employ anomalous $U(1)_{A}$ flavor symmetry (see e.g.~\cite{Ibanez:1994ig, Binetruy:1996xk, Dudas:1996fe,Grossman:1998jj,Dreiner:2003yr}) to address the `origin of fermion mass hierarchy and mixing and establish the desired flavor structure through the Froggatt-Nielsen (FN) mechanism~\cite{Froggatt:1978nt}.
The interplay between $U(1)_{A}$  and discrete symmetries play an essential role in suppressing dangerous proton decay operators in the MSSM. As we Further, as we will show, the size of the smallness of the Dirac neutrino masses is closely related to the mixed $U(1)_{A}$- gravitational anomaly cancellation.~Our model is compatible with $\SU{5}$ Grand Unification (GUT).~Since anomalous $\U{1}_{A}$ as well as discrete $R$ symmetries are expected to appear in string constructions, our setting is well motivated from a more fundamental theory.

This manuscript is structured as follows.
Section~\ref{sec:model} describes our $\SU{5}$-compatible model
based on a discrete $\Z{12}^R$ symmetry combined with a anomalous $\U1_{A}$ flavor symmetry. We discuss Green-Schwarz anomaly cancellation and review the general features of the model. In Section~\ref{sec:flavor_symmetry}, we promote $\U1_{A}$ to a full flavor symmetry.
We comment how modifying the charge assignment of the flavor model proposed in \cite{Babu:2003zz},
allows to cancel the gravitational anomaly without extra singlets and naturally explain the size of the Dirac neutrino masses in our model, in addition to explaining  the top-bottom hierarchy in the Yukawa sector. The textures we obtain are shown to be fully compatible with $\SU{5}$ and correctly reproduce the mixing in the quark and lepton sector, as shown by performing a $\chi^2$ fit.
In Section~\ref{sec:nucleon}, we discuss constraints from proton decay and comment on neutron--anti-neutron oscillations. We then demonstrate in Section~\ref{sec:cogenesis} that  the co-genesis of baryon number asymmetry and dark matter abundance is possible in our specific setting. Finally, Section~\ref{sec:conclusion} concludes the paper.

\section{Model}
\label{sec:model}

\subsection{Symmetries and particle content}
The model is based on a generation-independent $\Z{12}^R$ $R$-symmetry in combination of an anomalous $U(1)_{A}$ flavor symmetry. The particle content of the model with the charge assignments of the superfields is given in Table~\ref{fig:model}. From the charge assignment, it is evident that our model is $\SU{5}$-compatible with $\mathbf{10} = (Q, \overline{U}, \overline{E})$ and $\overline{\mathbf{5}} = (\overline{D}, L)$ 
being the $\SU{5}$ chiral GUT super-multiplets unifying the matter fields in each SM generation, while $\Phi$ and  $\overline{\Phi}$ being the $\SU{5}$ singlet chiral superfields. $\mathscr{W}_{hid}$ is the ``hidden''
sector superpotential and $X$, $S$ and $\sigma$ are the spurion fields parametrizing the SUSY/$R$ breaking, the dilaton and the flavon, respectively.
The ``$\rightarrow \Z{3}$" notation represents the residual $\Z{3}$ symmetry 
of $\Z{12}^R$ after SUSY/$R$ breaking.
\begin{figure}[htb]
\centering
  \setlength{\extrarowheight}{3pt}
  \setlength{\tabcolsep}{5pt}
\begin{tabular}{|c|cccccccc|cc|c|cc|c|}
\multicolumn{1}{c}{} &  \multicolumn{3}{c}{\begin{minipage}[c][12mm][c]{18mm}\begin{center}$\SU{5}$\\$\mathbf{10}$ \\ $\overbrace{\hspace*{15mm}}$\end{center} \end{minipage}} 
& \multicolumn{2}{c}{\begin{minipage}[c][12mm][c]{20mm}\begin{center}$\SU{5}$\\$\overline{\mathbf{5}}$ \\ $\overbrace{\hspace*{20mm}}$\end{center} \end{minipage}} 
&\multicolumn{6}{c}{}
& \multicolumn{1}{c}{\begin{minipage}[c][5mm][c]{10mm}\begin{center}$~$\\$\mathscr{W}_{\rm hid}$ \\ $\overbrace{\hspace{6mm}}$\end{center} \end{minipage}}
\\ 
\hline
 & $Q$ & $\overline{U}$ & $\overline{E}$ & $\overline{D}$ & $L$ & $H_u$ & $H_d$ & $\overline{\nu}$ &  $\Phi$ 
& $\overline{\Phi}$ &   $\sigma$ & $e^{-b S}$ & $X$ &   $\theta$\\
\hline
\hline
$\Z{12}^R$ & 2 & 2 & 2 & 6 & 6 & 2 & 10 & 10 & 4 & 8 &    0 & 6 & $r_X$ &   3\\ 
$\rightarrow \Z{3}$ & 2 & 2 & 2 & 0 & 0 & 1 & 2 & 1   &   2 &   1 & 0 & 0 & $r_{X^{\prime}}$ &   - \\ 
$\Z{2}^M$  & 1 & 1 & 1 & 1 & 1 & 0 & 0   & 1 & 1 & 1 &    0 & 0 & 0 &   0\\  
\hline 
\multirow{3}{*}{$\U{1}_{A}$}
& 3 & 3 & 3 & $(1+p)$ & $(1+p)$ & \multirow{3}{*}{0} & \multirow{3}{*}{0}   & $(15 + p)$ & \multirow{3}{*}{0} & \multirow{3}{*}{0} & \multirow{3}{*}{-1} & \multirow{3}{*}{$q_S$} & \multirow{3}{*}{0} & \multirow{3}{*}{0}   \\  
& 2 & 2 & 2 & $p$ & $p$ &   &     & $(14 + p)$ &   &   &      &   &      &  \\  
& 0 & 0 & 0 & $p$ & $p$ &   &    & $(13 + p)$ &   &   &      &   &   &    \\
\hline
\end{tabular}
\captionof{table}{The particle content and charge assignment of the model. All fields shown are chiral multiplets; $\theta$ denotes the super-space coordinates. The three horizontal rows at the bottom represent the $\U{1}_A$ charges for each of the fermions in the three families, respectively.}
\label{fig:model}
\end{figure}

The gauge invariant superpotential, up to order 4, is given by,
\begin{align} \label{eq:base_sup}
\mathscr{W} =~ & Y_e L \Hd \overline{E} + Y_d Q \Hd \overline{D} + Y_u Q \Hu \overline{U}  + Y_{\nu} L \Hu \overline{\nu} \nonumber\\
& + \kappa_1 \overline{U} \overline{D} \overline{D} \Phi + \kappa_2  L L \overline{E} \Phi 
+ \kappa_3  L Q \overline{D} \Phi  \; ,
\end{align}
where $Y_{u, \, d, \, e, \, \nu}$, implicitly depend on powers of the flavon field, $\sigma$. After the scalar component of the flavon field, $\sigma$, acquires a vacuum expectation value (VEV) breaking the $U(1)_{A}$ flavor  symmetry, the effective Yukawa couplings are generated, as discussed in Section~\ref{sec:u1A}. 
Note that
$\Phi^n$ and $\overline{\Phi}^n$ are forbidden in the superpotential to all orders in $n$. Additionally, there will be terms in the effective superpotential that appear ``non-perturbatively'' after SUSY breaking,
\begin{align} \label{eq:nonp_sup}
\mathscr{W}_{eff}^{np} \supset ~ & \mu \Hu \Hd + M_{\Phi} \Phi \overline{\Phi} 
+ \kappa_4 Q Q Q L + \kappa_5 \overline{U} \overline{U} \overline{D} \overline{E} \\
& + \kappa_6 \overline{U} \overline{D} \overline{D} \overline{\nu} 
+ \kappa_7  L L \overline{E}  \overline{\nu} 
+ \kappa_8  L Q \overline{D}  \overline{\nu} +\dots \nonumber
\end{align}
\noindent The order parameter for SUSY/$R$-symmetry breaking is the VEV of ``hidden sector'' superpotential 
$\langle \mathscr{W}_{hid} \rangle \sim m_{3/2}$ as in gravity mediation, which allows to parametrize the size of the effective terms. 

Some key features of the model are described in the following:
\begin{itemize}

\item {\bf$\mu$-term}: The $\mu$ term is forbidden in the superpotential due to the discrete R-symmetry, and appears only after SUSY and R-symmetry breaking. It can be generated with the correct magnitude through the Giudice-Masiero mechanism \cite{Giudice:1988yz}.
Specifically, the $\mu$ term can effectively appear 
from the K\"ahler potential  
\begin{equation}
K \supset k_{\Hu \Hd} \dfrac{X^{\dagger}}{M_{pl}} \Hu \Hd + h.c. 
\end{equation}
once the anti-holomorphic $X^{\dagger}$ field acquires a VEV in the $F$-term, $\langle F_{X} \rangle \sim m_{3/2} M_{pl}$, with $M_{pl} = 2.435 \times 10^{18}$ GeV being the reduced Planck mass. This leads to $\mu_{eff}$ to be of the order of the gravitino mass, as in gravity mediation. Note that in the model of \cite{Kitano:2008tk}, based on non-$R$ discrete symmetry, the $\mu$-problem persists. On the other hand, discrete $R$-symmetries allow to naturally address this issue~\cite{Chen:2012jg}.

\item {\bf Neutrinos:} The Weinberg operator, $L\Hu L \Hu$, which leads to Majorana neutrino masses, is forbidden in both the superpotential and the  K\"ahler-potential to all orders due to the discrete $R$ symmetry. The neutrinos are Dirac fermions. As we show below, the suppression in the Dirac  neutrino masses in our model naturally follow from the family $\U{1}_A$ symmetry, and its magnitude is determined by the requirement of cancellation of mixed $U(1)_{A}$-gravitational anomaly.

\item {\bf Flavor structure:} The flavor structure of the model is determined by the Froggatt-Nielsen mechanism \cite{Froggatt:1978nt}. This approach has already been extensively studied in the literature (e.g. \cite{Ibanez:1994ig,Binetruy:1996xk,Dudas:1996fe,Grossman:1998jj,Dreiner:2003yr}).~Here, the Yukawa couplings are assumed to be of order one, with the specific textures and
hierarchy determined by the powers of a small expansion parameter $\varepsilon$.
More specifically, the SM fermions carry various positive charges under the $\U{1}_A$ symmetry, while the MSSM singlet flavon field $\sigma$ is negatively charged. As the flavon field acquires a VEV $\langle \sigma \rangle$ and breaks $\U{1}_A$, the Yukawa terms obtain $(Y \langle \sigma \rangle/ M)^n$ couplings, where $Y \simeq \mathcal{O}(1)$, $\epsilon  = (\langle \sigma \rangle/M)$ and $M$ is some suppression scale.

\item {\bf Proton decay:} The potentially dangerous proton decay operators, such as $QQQL$, do not appear in the superpotential, due to the discrete $R$-symmetry (see e.g. \cite{Chen:2014gua}). The ``non-perturbative'' contribution from the K\"ahler potential is insignificant.

\item {\bf Baryogenesis:} 
The  chiral superfield $\Phi = (\phi, \tilde{\phi}, F_{\phi})$ couples to baryon number-carrying operator $\overline{U} \overline{D} \overline{D}$  and appears as a higher-demensional superpotential term 
\begin{equation} \label{eq:baropp}
\mathscr{W} \supset \dfrac{1}{M} \Phi\overline{U}\overline{D}\overline{D}~.
\end{equation}
This term will be responsible for generating the baryon asymmetry in the model \'a la Affleck-Dine mechanism \cite{Affleck:1984fy}, as follows. First, an asymmetry in $\phi$ is generated. Coherent oscillations of the $\phi$-field come to dominate the energy density of the universe after inflation, on-setting matter-domination era. As $\phi$ later decays, prior to the BBN, the universe is reheated and the $\phi$-asymmetry is transferred to the baryons through the operator in Eq.~\eqref{eq:baropp}. Our treatment of baryogenesis closely follows that of \cite{Kitano:2008tk} (see \cite{Dimopoulos:1987rk,Thomas:1995ze,Cline:1990bw,Fujii:2002aj,Moroi:1999zb} for earlier similar works).
\end{itemize}

\subsection{Green-Schwarz anomaly cancellation}

\subsubsection{$\U{1}_A$ anomalies}

The gauge anomalies associated with $\U{1}_{A}$ can be canceled via the four-dimensional Green-Schwarz mechanism as follows. 
The scalar component of the dilaton chiral superfield $S$, $S|_{\theta = 0} = s + i a$, contains the dilaton field $s$ as well as the string axion field $a$. In the presence of $S$, the gauge boson Lagrangian is given by
\begin{equation}
\mathscr{L}_{\rm gauge} = \dfrac{1}{4} \int d^2\theta~\Big[ k_{A} S \Tr{W_{A}^{\alpha} W_{{A} \alpha}} + k_{\rm a} S \Tr{W_{\rm a}^{\alpha} W_{{\rm a} \alpha}} \Big] + \rm{h.c.}~,
\end{equation}
where $W_{{A}}^{\alpha}$, $W_{a}^{\alpha}$ are the superfield strengths of vector supermultiplets $V_{A}$, $V_a$, and $k_{A}$, $k_a$ are the Kac-Moody algebra levels, associated with the gauge groups $\U{1}_{A}$ and $G_a$, respectively. For non-Abelian groups, Kac-Moody levels are positive integers, while for $\U{1}$ they need not be integers. The gauge couplings are related to the dilaton VEV as $k_a \langle S \rangle = 1/g_a^2$. The above leads to the axion-coupled field strength Lagrangian terms
\begin{equation}  \label{eq:saxlag}
\mathscr{L} \supset - \dfrac{a}{8} F^{A} \tilde{F}^{A} - \dfrac{a}{8} F^{a} \tilde{F}^{a} + \dfrac{a}{4} \mathcal{R} \tilde{\mathcal{R}}~,
\end{equation}
where $F^{A}$, $F^a$ denote the gauge field strength of $\U{1}_{A}$, $G_a$ and $\mathcal{R}$ is the Riemann curvature tensor, with the last term describing the gravity contribution.

Under the $\U{1}_{A}$ gauge transformations the dilaton $S$ shifts as  
\begin{equation} \label{eq:shift}
S \rightarrow S + \dfrac{i}{2} \delta_{\rm GS} \Lambda~,
\end{equation}
where $\Lambda$ is a parameter chiral superfield and $\delta_{\rm GS}$ is a real number. The associated vector superfield transforms as $V_A \rightarrow V_A + (i/2) (\Lambda - \Lambda^{\dagger})$. Hence, gauge invariance requires the modified K\"ahler potential of the dilaton to be of the form $K(S, S^{\dagger}, V_A) = - \log(S + S^{\dagger} - \delta_{\rm GS} V_A)$.

Chiral anomalies contribute terms of the form $F \tilde{F} \mathcal{A}$, where $\mathcal{A}$ is the anomaly coefficient and $F$ is the field strength of some gauge group. The gravitational anomaly contribution is similar. As the dilaton transformations of Eq.~\eqref{eq:shift} induce an axion shift in terms of Eq.~\eqref{eq:saxlag}, the anomaly contributions, which enter with the opposite sign, are canceled by an appropriate choice of $\delta_{\rm GS}$. 
The requirement of anomaly cancellation is fulfilled by (e.g. \cite{Maekawa:2001uk})
\begin{equation} \label{eq:gscondition}
\dfrac{\mathcal{A}_{CCA}}{k_{C}} = \dfrac{\mathcal{A}_{WWA}}{k_{W}} = \dfrac{\mathcal{A}_{YYA}}{k_{Y}} =
\dfrac{\mathcal{A}_{AAA}}{3 k_{A}} = 
\dfrac{\mathcal{A}_{GGA}}{24} = 
2 \pi^2 \delta_{\rm GS}~,
\end{equation}
where $\mathcal{A}_{\cdots}$ denote coefficients from the $\big[\SU{3}_{C}\big]^2 \times \U{1}_{A}$, $\big[\SU{2}_{W}\big]^2 \times \U{1}_{A}$, $\big[\U{1}_{Y}\big]^2 \times \U{1}_{A}$, $\big[\U{1}_{A}\big]^3$ and the  $\big[\text{gravity}\big]^2 \times \U{1}_{A}$ anomalies\footnote{Factor of 3 for $\big[\U{1}_{A}\big]^3$ is combinatorial, due to the anomaly being pure and originating from identical groups.}. Since the trace of $\SU{N}$ generators vanishes, cross anomalies such as $\big[\SU{3}_{C}\big] \times \U{1}_{A}^2$ are automatically zero. The anomaly coefficient $\mathcal{A}_{YAA}$ for $\big[\U{1}_{A}\big]^2 \times \U{1}_{Y}$ should also vanish. The anomaly coefficients are given as follows (see, for example, \cite{Dreiner:2003yr})
\begin{equation} \label{eq:anom_u1}
\left\{\begin{array}{lll}
\mathcal{A}_{CCA} & =~\dfrac{1}{2} \Big[ \sum\limits_{i = 1}^3   (3   q_{\bf{10}}^i + q_{\overline{\bf{5}}}^i) \Big]  \\
\mathcal{A}_{WWA} & =~\dfrac{1}{2} \Big[ (q_{\Hu} + q_{\Hd}) + \sum\limits_{i = 1}^3  (3   q_{\bf{10}}^i + q_{\overline{\bf{5}}}^i) \Big]  \\
\mathcal{A}_{YYA} & =~\dfrac{1}{2} \Big[ (q_{\Hu} + q_{\Hd}) + \dfrac{5}{3} \sum\limits_{i = 1}^3 (3   q_{\bf{10}}^i +  q_{\overline{\bf{5}}}^i) \Big] \cdot \dfrac{3}{5} \\
\mathcal{A}_{AAA} & =~2   (q_{\Hu}^3 + q_{\Hd}^3) +  5 \sum\limits_{i = 1}^3 (2  (q_{\bf{10}}^i)^3 +  (q_{\overline{\bf{5}}}^i)^3)   + q_{\sigma}^3 + \sum\limits_{i = 1}^3 (q_{\overline{N}}^i)^3 + \mathcal{A}_{YAA}^{\rm{hidden}}
\\
A_{GGA} & =~2   (q_{\Hu}  + q_{\Hd} ) +  5 \sum\limits_{i = 1}^3 (2   q_{\bf{10}}^i  +   q_{\overline{\bf{5}}}^i )   + \sum\limits_{\rm{SM~singlet}} q_s   + \mathcal{A}_{GGA}^{\rm{hidden}} \\
\end{array}\right.
\end{equation}
where we have already accounted for $\SU{5}$ charge multiplet assignment and the standard normalization of $1/2$ for the  fundamental representation of $\SU{N}$. The hypercharge $Y_L$ is not quantized in general and the uncertainty in its normalization renders the anomaly coefficients $ \mathcal{A}_{YAA}$ not very informative (see discussion in \cite{Dreiner:2005rd}). The additional possible anomaly coefficient $\mathcal{A}_{YAA}$ due to $\big[\U{1}_{A}\big]^2 \times \U{1}_{Y}$ also suffers from the hypercharge normalization uncertainty and is often neglected. For $\SU{5}$ matter charge assignment as well as $q_{\Hu} = q_{\Hd}$, which is the case of our model, it vanishes. The hypercharge normalization can be fixed by an underlying GUT. The pure $[\U{1}_A]^3$ and gravitational anomalies can obtain potential contributions from the hidden sector SM singlet fields and are thus often neglected. However, the existence of additional light
singlet states is potentially of phenomenological interest and gravitational anomaly could still be useful. Thus, we shall only focus on $\mathcal{A}_{CCA}$, $\mathcal{A}_{WWA}$, $\mathcal{A}_{YYA}$ and $\mathcal{A}_{GGA}$ as has been typically done in previous similar studies (e.g. \cite{Babu:2003zz}). We note that since SM singlet fields do not carry a non-trivial Dynkin index they only contribute to the gravitational anomaly coefficient.

Assuming gauge-coupling unification, which can occur without a simple covering group as in string theory, one has  \cite{Ginsparg:1987ee}
\begin{equation}
g_C^2 k_C = g_W^2 k_W = g_Y^2 k_Y = g_A^2 k_A = 2 g_s^2~,
\end{equation}
where $g_s$ is the string coupling constant. Enforcing the observed gauge-coupling unification in the MSSM  requires 
\begin{equation}
g_C = g_W = \sqrt{\dfrac{5}{3}}  g_Y~,
\end{equation}
resulting in
\begin{equation}
k_C = k_W =  \dfrac{3}{5}  k_Y~.
\end{equation}
This is the same as $\SU{5}$ hypercharge normalization. 
The Green-Schwarz anomaly cancellation condition \eqref{eq:gscondition} can be satisfied for the $\U{1}_A$ charge assignment of Table~\ref{fig:model} by 
considering higher Kac-Moody levels\footnote{Non-minimal Kac-Moody levels also appear in string
theory and can give rise to interesting phenomenological features \cite{Aldazabal:1994zk,Dienes:1995sq}} 
\begin{equation}
 k_C = k_W = (5/3) k_Y = 2~, 
\end{equation} 
with
\begin{equation} \label{eq:anomu1a}
\dfrac{\mathcal{A}_{CCA}}{k_{C}} = \dfrac{\mathcal{A}_{WWA}}{k_{W}} = \dfrac{\mathcal{A}_{YYA}}{k_{Y}} =
\dfrac{\mathcal{A}_{GGA}}{24} =  \dfrac{(16 + 3p)}{4}~.
\end{equation}

\subsubsection{Discrete $R$-symmetry anomalies}

For a local $\U{1}$ $R$-symmetry, the superspace variable $\theta$ carries a charge of $q_{\theta} = R$ and hence the superpotential carries $2 q_{\theta}$, with $R$ being an integer. For a chiral superfield carrying a charge of $q_x$, the associated fermion carries a charge of $(q_x - q_{\theta})$. This is also the case for Higgsinos, while gauginos carry a charge of $q_{\theta}$. The corresponding anomaly coefficients are then  \cite{Lee:2011dya, Dreiner:2012ae}
\begin{equation} \label{eq:anom_u1r}
\left\{\begin{array}{lll}
\mathcal{A}_{CCR} & =~\dfrac{1}{2} \Big[ \sum\limits_{i = 1}^3   (3   q_{\bf{10}}^i + q_{\overline{\bf{5}}}^i) \Big] - 3 R \\
\mathcal{A}_{WWR} & =~\dfrac{1}{2} \Big[ (q_{\Hu} + q_{\Hd}) + \sum\limits_{i = 1}^3  (3   q_{\bf{10}}^i + q_{\overline{\bf{5}}}^i) \Big] - 5 R \\
\mathcal{A}_{YYR} & =~\dfrac{1}{2} \Big[ (q_{\Hu} + q_{\Hd} - 11 R) + \dfrac{5}{3} \sum\limits_{i = 1}^3 (3   q_{\bf{10}}^i +  q_{\overline{\bf{5}}}^i) \Big] \cdot \dfrac{3}{5} 
\\
\mathcal{A}_{GGR} & =~2   (q_{\Hu}  + q_{\Hd} - 2 R ) +  5 \sum\limits_{i = 1}^3 (2   q_{\bf{10}}^i  +   q_{\overline{\bf{5}}}^i - 3 R)   + \sum\limits_{\rm{SM~singlet}} q_s  \\
& ~~~ + \mathcal{A}_{GGR}^{\rm{hidden}} + 33 R\\
\end{array}\right.
\end{equation}

For a gauged discrete $\Z{N}$ symmetry anomalies should also cancel. While one cannot construct the corresponding triangle diagram, as there is no gauge boson associated with the discrete symmetry, the anomaly coefficient can still be computed via the Fujikawa method using the path integral measure transformation. The anomaly cancellation condition now becomes 
\begin{equation}
\dfrac{\mathcal{A}_{CCZ}}{k_{C}} = \dfrac{\mathcal{A}_{WWZ}}{k_{W}} = \dfrac{\mathcal{A}_{YYZ}}{k_{Y}} =
\dfrac{\mathcal{A}_{GGZ}}{24} =  \pi N \Delta_{\rm GS}\mod{\eta}~,
\end{equation}
where $\Delta_{\rm GS}$ is the dilaton shift constant, and 
\begin{equation}
  \eta=\begin{cases}
    N, & \text{for $N$ odd} \\
    N/2, & \text{for $N$ even} ~.
  \end{cases}
\end{equation}
In the case of discrete $R$ symmetry, $\Z{N}^R$, the computation is similar. For $\Z{12}^R$ and charges of Table~\ref{fig:model} the corresponding anomaly coefficients are
\begin{equation} 
\left\{\begin{array}{lll}
\mathcal{A}_{CCR} & =~(9 \cdot k_{C}) \mod{6} ~=~3\\
\mathcal{A}_{WWR} & =~(9 \cdot k_{W})\mod{6}~=~3 \\
\mathcal{A}_{YYR} & =~(\dfrac{9}{5} \cdot k_{Y})\mod{6}~=~3 \\
\mathcal{A}_{GGR} & =~(42 \cdot 24)\mod{6}~=~0 \\
\end{array}\right.
\end{equation}
where $k_{C} = k_{W} = (5/3) k_{Y} = 1$ (all of the anomaly coefficients are 0 for $k_{C} = k_{W} = (5/3) k_{Y} = 2$).

Since in our setup both $\U{1}_A$ as well as discrete gauge symmetries are present, mixed anomaly coefficients involving both of these can appear.
These coefficients are usually neglected, as they depend on the normalization of the $\U{1}_A$ charges \cite{Ibanez:1992ji} and we will ignore them.

\section{Flavor structure and neutrino masses}
\label{sec:flavor_symmetry}

\subsection{Anomalous $\U{1}_A$ flavor symmetry}
\label{sec:u1A}

The SM Cabibbo-Kobayashi-Maskawa (CKM) quark matrix can be parametrized through the Cabbibo angle $\sin \theta_c \simeq 0.2$. The Yukawa coupling suppression $(\langle \sigma \rangle/M)$ naturally provides the desired small expansion parameter $\epsilon$ for the flavor structure of the model of the size of the Cabbibo angle, assuming that the model is string-inspired and identifying the scale $M$ with the string scale $M_{s}$. For simplicity, we take $M_{s} = M_{pl}$ throughout this work. Anomalous flavor $\U{1}_A$ can also be used to forbid certain operators, through supersymmetric zeroes\footnote{In this case, the flavon coupling is family-independent and matter fields are allowed to carry negative charges under $\U{1}_A$. Since holomorphicity of superpotential only allows for $\sigma$ and not $\sigma^{\dagger}$ to appear, any combined operator that carries a total negative $\U{1}_A$ charge will be forbidden.} (e.g. \cite{Leurer:1992wg}).

The modified K\"ahler potential invariant under the dilaton transformations \eqref{eq:shift} leads to a radiatively-generated string Fayet-Iliolopuolos (FI) $\U{1}_A$ term from the gravitational anomaly \cite{Dine:1986zy,Dine:1987bq,Dine:1987gj,Atick:1987gy}, with a coefficient 
\begin{equation}
\xi_{\rm FI} = \dfrac{g_s^2}{192 \pi^2} M_{pl}^2 A_{GGA}~,
\end{equation}
where $A_{GGA}$ is the gravitational anomaly coefficient as before. Inclusion of a non-zero FI term leads to a D-term contribution to the scalar potential of the form
\begin{equation}
V_{\rm scalar} \supset \dfrac{1}{2} D^2 = \dfrac{g_A^2}{2} \Big( \sum_i q_{\phi_i} |\phi_i|^2 + \xi_{\rm FI}^2 \Big)^2~,
\end{equation}
where $g_A$ is the $\U{1}_A$ gauge coupling as before and $\phi_i$ is the scalar component of the $\Phi_i$ chiral superfield carrying a $\U{1}_A$ charge of $q_{\phi_i}$. Requirement of $\SU{3}_C \times \SU{2}_W \times \U{1}_Y$ gauge invariance ensures that the scalar components of MSSM-superfields have vanishing VEVs. In order to maintain SUSY at the $\U{1}_A$-breaking scale, the D-term must vanish. This requires that at least one remaining chiral superfield has a negative charge to cancel the $\xi_{\rm FI}$ contribution\footnote{A $\U{1}_A$ charge normalization resulting in $q_{\sigma} = -1$ is assumed.}. This role is played by the flavon field $\sigma$, setting its VEV to $\langle \sigma \rangle \propto \sqrt{\xi_{\rm FI}}$. The previously obtained gravitational anomaly coefficient in Eq.~\eqref{eq:anomu1a} of $(16 + 3p)/4 $, with $p \in \{0,1,2\}$, thus leads to a desired value of the small expansion parameter
\begin{equation}
 \epsilon = \dfrac{\langle \sigma \rangle}{M_{pl}} = \sqrt{\dfrac{g_s^2 A_{\rm GGA}}{192 \pi^2}} \simeq \sin \theta_c = \mathcal{O}(0.2)~, 
 \end{equation}
 where we have taken $g_s^2/4 \pi \simeq 1/24$.
 
After $\U{1}_A$ breaking the corresponding flavor gauge boson and the gaugino become massive, with a mass of
\begin{equation}
M_A = \dfrac{g_A \langle \sigma \rangle}{\sqrt{2}} \simeq \mathcal{O}(10^{-2}) M_{pl}~.
\end{equation}
Between the scale $M_{pl}$ and $M_A$ they can contribute to flavor violating processes, with potentially observable consequences \cite{Babu:2004th}.

\subsection{Textures and massive Dirac neutrinos}

With the flavon field, the Yukawa terms appear in the superpotential as 
\begin{equation}
\begin{array}{ll}
\mathscr{W} \supset & y_{ij}^e \Big(\dfrac{\sigma}{M_{pl}}\Big)^{n_{ij}^e} \, L_i \, \Hd \, \overline{E}_j +
 y_{ij}^d \Big(\dfrac{\sigma}{M_{pl}}\Big)^{n_{ij}^d} \, Q_i \, \Hd \, \overline{D}_j \\
& +~y_{ij}^u \Big(\dfrac{\sigma}{M_{pl}}\Big)^{n_{ij}^u} \, Q_i \, \Hu \, \overline{U}_j +
y_{ij}^{\nu} \Big(\dfrac{\sigma}{M_{pl}}\Big)^{n_{ij}^{\nu}} \, L_i \, \Hu \, \overline{\nu}_j~,
\end{array}
\end{equation}
where $i,j$ denote the flavor indices. Here,
$y_{ij}^f$ are the Yukawa coupling coefficients of $\mathcal{O}(1)$
and $n_{ij}^f$ are the positive integers determined by the $\U1_{A}$ charge assignment for the fermions in family $f$.
After flavor symmetry breaking, the Yukawa couplings take the form of
\begin{equation}
Y_{ij}^f = y_{ij}^f \Big(\dfrac{\sigma}{M_{pl}}\Big)^{n_{ij}^f}
 \longrightarrow ~y_{ij}^f \Big(\dfrac{\langle\sigma\rangle}{M_{pl}}\Big)^{n_{ij}^f}
 = y_{ij}^f \epsilon^{n_{ij}^f}~.
\end{equation}

For the fermion $\U{1}_A$ charge assignment of Table~\ref{fig:model} as well as Yukawa coupling parametrization, neglecting right handed neutrinos, we employed the results of Model 2 (i.e. $\alpha = 1$) of~\cite{Babu:2003zz}. These $\SU{5}$-compatible flavor textures are a variation
of textures presented in \cite{Babu:2002tx}.~The different values of parameter $p \in \{2,1,0\}$ correspond to different values of
 $\tan \beta \equiv \langle\Hu\rangle / \langle\Hd\rangle \in \{5, 10, 20\}$,
respectively.
These charges give the following mass matrices 
\[ \label{eq:yukawa_form}
M_u \sim \langle \Hu \rangle \left(\begin{array}{@{}ccc@{}}
     \epsilon^{6} &  \epsilon^{5} & \epsilon^{3} \\
     \epsilon^{5} &   \epsilon^{4} & \epsilon^{2} \\
      \epsilon^{3} &  \epsilon^{2} & 1 \\
\end{array}\right)
~~~;~~~
M_d \sim \langle \Hd \rangle \epsilon^p \left(\begin{array}{@{}ccc@{}}
     \epsilon^{4} &  \epsilon^{3} & \epsilon^{3} \\
     \epsilon^{3} &   \epsilon^{2} & \epsilon^{2} \\
      \epsilon & 1 & 1 \\
\end{array}\right)
~~~;~~~
M_e \sim \langle \Hd \rangle \epsilon^p \left(\begin{array}{@{}ccc@{}}
     \epsilon^{4} &  \epsilon^{3} & \epsilon \\
     \epsilon^{3} &   \epsilon^{2} & 1 \\
      \epsilon^{3} &  \epsilon^{2} & 1 \\
\end{array}\right)~~~,
\]
yielding approximately the following mass relations
\begin{eqnarray}
m_u : m_c : m_t ~\sim & \epsilon^6 : \epsilon^4 : 1 \nonumber\\
m_d : m_s : m_b ~\sim  & \epsilon^4 : \epsilon^2 : 1 \\
m_e : m_{\mu} : m_{\tau} ~\sim  & \epsilon^4 : \epsilon^2 : 1  \nonumber
\end{eqnarray}
Analysis of renormalization group equations (RGEs) as well as arbitrary $\mathcal{O}(1)$ coefficients in front
of the entries have been also provided by \cite{Babu:2003zz}.~Using \texttt{REAP/MPT} packages \cite{Antusch:2005gp}, we have confirmed that these textures yield approximately correct~mixings.
 
From $\SU{5}$ compatibility, $Y_e = Y_d^T$.
However, this leads to a known issue of unacceptable fermion mass relations, with $m_d/m_e = m_s/m_{\mu} = m_b/m_{\tau}$. A possible solution has been proposed by Georgi and Jarlskog \cite{Georgi:1979df}, which introduced additional Higgs multiplets that don't couple to all the generations. Adopting their approach, we modify $M_e$ as follows
\[ \label{eq:modye}
M_e \sim \langle \Hd \rangle \epsilon^p \left(\begin{array}{@{}ccc@{}}
     \epsilon^{4} &  \epsilon^{3} & \epsilon \\
     \epsilon^{3} &   \epsilon^{2} & 1 \\
      \epsilon^{3} &  -3 \epsilon^{2} & 1 \\
\end{array}\right)~~~.
\]

For the neutrinos, in contrast to \cite{Babu:2003zz}, we do not rely on the see-saw mechanism for generation of neutrino masses. Neutrinos in our model are thus Dirac fermions. Without the see-saw mechanism in play, the smallness of the neutrino masses begs for an explanation. We achieve this by requiring that the mixed $\U{1}_A$-gravitational anomaly  \eqref{eq:anomu1a} is canceled by the GS mechanism without the need of invoking any extra singlet fields, in contrast to \cite{Babu:2003zz}. This condition  sets the right-handed neutrino charges to the values presented in Table~\ref{fig:model} and automatically leads to the desired neutrino mass suppression through the Froggatt-Nielson expansion parameter $\epsilon$. The resulting modified Dirac neutrino Yukawa texture is 
\[
M_{\nu} \sim \langle \Hu \rangle \epsilon^{13 + 2p} \left(\begin{array}{@{}ccc@{}}
     \epsilon^{3} &  \epsilon^2  & \epsilon  \\
     \epsilon^2  &  \epsilon & 1 \\
      \epsilon^2 & \epsilon & 1 \\
\end{array}\right)~,
\]
which gives
reasonable neutrino mixing, as demonstrated below.
 For different values of $p$, one obtains the desired neutrino mass suppression of
\begin{equation}
\begin{array}{ll}
p \in \{0, 1, 2\} \longrightarrow \dfrac{m_{\nu}}{\langle H_u \rangle} \sim  
\{\epsilon^{13}, \epsilon^{14}, \epsilon^{15}  \} \simeq 
\{ 10^{-9}, 10^{-11}, 10^{-12} \}~.
 \end{array}
\end{equation} 

\subsection{Fitting fermion masses and mixings}

We demonstrate that the above Yukawa textures
can lead to reasonable fermion mixings and mass hierarchies by explicitly fitting the $\mathcal{O}(1)$ coefficients in front of $\epsilon$ factors.
As the $\sigma$ flavon field obtains a VEV at the $\U1_{A}$ breaking scale of $\sim 10^{15}$ GeV, 
the resulting Yukawa matrices should in principle be evolved with RGEs down to the scale $M_Z$, where the parameters are to be compared to the measured experimental values.
Since our goal here is only to
show that we can obtain approximately correct flavor structure
at the GUT scale, we will not discuss the RGE analysis (see e.g. \cite{Babu:2003zz} for a possible implementation).

The Yukawa matrices are diagonalized using singular value decomposition (SVD), which explicitly gives the left- and right-handed unitary rotation matrices $V_i$. The respective rotation matrices for the quark and lepton fields are 
\begin{align}
d_R~, u_R~, Q_L & \rightarrow V_R^d d_R~, V_R^u u_R~, V_L^Q Q_L \\
e_R~, \nu_R~, L_L & \rightarrow (V_R^e)^{\dagger} e_R~,
 (V_R^{\nu})^{\dagger} \nu_R ~, (V_L^{L})^{\dagger} L_L~.  \notag
\end{align}
The CKM and PMNS mixing matrices are then\footnote{The convention for the PMNS and CKM matrices 
in the weak currents is
$\mathscr{L}_{CC}^{quark} = \dfrac{g}{\sqrt{2}} \overline{u}_L \gamma^{\mu} W_{\mu}^{+} V_{\rm{CKM}} d_L$~+~h.c. 
for quarks and $\mathscr{L}_{CC}^{lep} =  \dfrac{g}{\sqrt{2}} \overline{e}_L \gamma^{\mu} W_{\mu}^{-} V_{\rm{PMNS}} \nu_L$~+~h.c 
for leptons.}
\begin{align}
V_{\rm{CKM}} & = (V_L^u)^{\dagger} V_L^d \\
V_{\rm{PMNS}} & = V_L^{e} (V_L^\nu)^{\dagger}~. \notag
\end{align}

The current measured values of the CKM mixing angles (see PDG \cite{Patrignani:2016xqp}), denoted by $q$, are
\begin{align}
(\theta_{12}^q, \theta_{13}^q, \theta_{23}^q, \delta_{\rm{CP}}^q) 
&= (12.98^{+0.026}_{-0.026},~0.204^{+0.015}_{-0.015},~2.42^{+0.053}_{-0.053},~71.03^{+1.91}_{-1.91})~. 
\end{align}
Similarly, the PMNS mixing angles \cite{Esteban:2016qun}, denoted by $l$, are 
\begin{align}
(\theta_{12}^l, \theta_{13}^l, \theta_{23}^l, \delta_{\rm CP}^l) 
&= (33.56^{+0.075}_{-0.075} ,~8.46^{+0.15}_{-0.15},~41.6^{+1.5}_{-1.5},~261^{+59}_{-59})~,
\end{align}
where we took the higher of the upper/lower uncertainty and have assumed for illustration purposes the normal hierarchy ordering, which is currently slightly favored by experiments.

In addition to the above mixing, if the model is to be compatible with $\SU{5}$,
at the GUT scale or near it\footnote{Since RGE running is logarithmic,
Yukawa coefficients at the $\U1_{A}$ breaking scale of $\sim 10^{15}$ GeV
would still strongly resemble their original GUT-scale relations, if there is an underlying
unifying group.}
quark and fermion masses should be related. As we have adopted the approach by Georgi and Jarlskog \cite{Georgi:1979df} and modified the $Y_e$ texture, the resulting GUT-scale masses should be related as 
\begin{equation}
3 m_e/m_d = m_{\mu}/3 m_s = m_{\tau}/ m_b~.
\end{equation}
This will ensure that at low energies the observed mass relations   approximately hold. Further, compatibility with $\SU{5}$ requires that not only do the textures obey $Y_d = Y_e^T$, but the respective $\mathcal{O}(1)$ coefficients are also related. We note that coefficients presented in the models of \cite{Babu:2003zz} are unrelated and thus, strictly speaking, their resulting flavor structures are not compatible with $\SU{5}$.

To find the relevant $\mathcal{O}(1)$ coefficients for the Yukawa matrices, we have generated 500,000 realizations of $Y_e, Y_d, Y_u, Y_{\nu}$. Each realization includes randomized complex $\mathcal{O}(1)$ coefficients in front of $\epsilon$'s.
From each realization we obtain the CKM and PMNS mixings as well as fermion masses with the help of the \texttt{MPT} package~\cite{Antusch:2005gp}.

To select the best fitting realization, we define a $\chi^2$-like statistic
\begin{equation}
\chi_{\rm tot}^2 = \chi_{\rm CKM}^2 + \chi_{\rm PMNS}^2 + \chi_{\rm mass}^2~,
\end{equation}
where the PMNS (denoted ``$l$'') and CKM (denoted ``$q$'') components are given as
\begin{align}
\chi_{\rm CKM}^2 = \dfrac{(\theta_{13}^{q,th} - \theta_{13}^{q,ex})^2}{(\sigma_{13}^{q,ex})^2}
+ \dfrac{(\theta_{23}^{q,th} - \theta_{23}^{q,ex})^2}{(\sigma_{23}^{q,ex})^2}
+ \dfrac{(\theta_{12}^{q,th} - \theta_{12}^{q,ex})^2}{(\sigma_{12}^{q,ex})^2}
+ \dfrac{(\delta_{\rm CP}^{q,th} - \delta_{\rm CP}^{q,ex})^2}{(\sigma_{\rm CP}^{q,ex})^2} \\
\chi_{\rm PMNS}^2 = \dfrac{(\theta_{13}^{l,th} - \theta_{13}^{l,ex})^2}{(\sigma_{13}^{l,ex})^2}
+ \dfrac{(\theta_{23}^{l,th} - \theta_{23}^{l,ex})^2}{(\sigma_{23}^{l, ex})^2}
+ \dfrac{(\theta_{12}^{l,th} - \theta_{12}^{l,ex})^2}{(\sigma_{12}^{l,ex})^2}
+ \dfrac{(\delta_{\rm CP}^{l,th} - \delta_{\rm CP}^{l,ex})^2}{(\sigma_{\rm CP}^{l,ex})^2}
\end{align}
Here, superscript ``$th$'' represents the theoretical values obtained from one realization, while ``$ex$'' denotes the experimentally observed values as well as their respective uncertainties.
The mass $\chi_{\rm tot}^2$ component
\begin{align}
\chi_{\rm mass}^2 = \dfrac{(m_e - m_d/3)^2}{m_e^2} + 
\dfrac{(m_{\mu} - 3 m_s)^2}{m_{\mu}^2} +
\dfrac{(m_{\tau} - m_{b})^2}{m_{\tau}^2}
\end{align}
enforces the Georgi-Jarlskog fermion mass relations. 

For $p = 0$, the best fit coefficients are found to be
\begin{align}
Y_u  &~=~~~~ \left(\begin{array}{@{}ccc@{}}
     ~[0.083 + 0.583 i]~ \epsilon^{6}~ &  ~[0.140 + 0.931 i]~ \epsilon^{5}~ & ~[0.121 + 0.572 i]~ \epsilon^{3}~ \\
     ~[0.935 + 0.962 i]~ \epsilon^{5}~ &  
      ~[0.233 + 0.302 i]~ \epsilon^{4}~ & 
       ~[0.454 + 0.467 i]~ \epsilon^{2}~ \\
      ~[0.272 + 0.388 i]~ \epsilon^{3}~ &  
      ~[0.129 + 0.612 i]~\epsilon^{2}~ & 
      ~[0.023 + 0.483 i]~~~~ \\
\end{array}\right) \\
Y_d  &~=~~~~ \left(\begin{array}{@{}ccc@{}}
    ~[0.361 + 0.900 i]~ \epsilon^{4} &  ~[0.377 + 0.689 i]~ \epsilon^{3} &  ~[0.906 + 0.470 i]~ \epsilon^{3} \\
     ~[0.058 + 0.739 i]~ \epsilon^{3} &   ~[0.789 + 0.223 i]~ \epsilon^{2} &   ~[0.217 + 0.665 i]~ \epsilon^{2} \\
     ~[0.619 + 0.298 i]~ \epsilon & ~[0.647 + 0.497 i]~~~ & ~[0.686 + 0.751 i]~~~ \\
\end{array}\right) \\
Y_e  &~=~~~~ \left(\begin{array}{@{}ccc@{}}
    ~[0.361 + 0.900 i]~ \epsilon^{4} &  ~[0.058 + 0.739 i]~ \epsilon^{3} & ~[0.619 + 0.298 i]~ \epsilon \\
     ~[0.377 + 0.689 i]~ \epsilon^{3} & ~[0.789 + 0.223 i]~  \epsilon^{2} & ~[0.647 + 0.497 i]~~~ \\
      ~[0.906 + 0.471 i]~ \epsilon^{3} &  ~[-0.651 - 1.995 i]~ \epsilon^{2} & ~[0.686 + 0.751 i]~~~ \\
\end{array}\right) \\
Y_{\nu}  &~=~ \epsilon^{13} \left(\begin{array}{@{}ccc@{}}
     ~[0.655 + 0.814 i]~ \epsilon^{3} &  ~[0.077 + 0.251 i]~ \epsilon^2  & ~[0.861 + 0.860 i]~ \epsilon  \\
     ~[0.166 + 0.724 i]~ \epsilon^2  &  ~[0.230 + 0.065 i]~ \epsilon & ~[0.291 + 0.205 i]~~~ \\
      ~[0.816 + 0.651 i]~ \epsilon^2 & ~[0.548 + 0.434 i]~ \epsilon & ~[0.637 + 0.232 i]~~~ \\
\end{array}\right) 
\end{align}
The textures above yield the following mixings
\begin{align}
(\theta_{12}^q, \theta_{13}^q, \theta_{23}^q, \delta_{\rm{CP}}^q) 
&\simeq (13.0^{\circ}, 0.3^{\circ},  2.2^{\circ},  65.4^{\circ}) \\
(\theta_{12}^l, \theta_{13}^l, \theta_{23}^l, \delta_{\rm{CP}}^l) 
&\simeq ( 34.6^{\circ},  14.2^{\circ},  46.9^{\circ},  181.4^{\circ}) \notag
\end{align}
as well as mass relations
\begin{equation}
\dfrac{3 m_e}{m_d} \simeq 1.3~~~,~~~ \dfrac{m_{\mu}}{3 m_s} \simeq 1.3 ~~~,~~~ \dfrac{m_{\tau}}{m_b} \simeq 1~.
\end{equation}
A better fit accuracy can be obtained with increasing the number of simulated realizations.

\section{Nucleon stability}
\label{sec:nucleon}

The $Q Q Q L$ and $\overline{U} \overline{U} \overline{D} \overline{E}$  baryon and lepton number violating operators of Eq.~\eqref{eq:nonp_sup} can
directly mediate proton decay via $p \rightarrow  \overline{\nu}K^+$, 
which is strongly constrained by the experiment \cite{Abe:2014mwa}.
Since these terms appear only  non-perturbatively 
in the effective superpotential, 
they are suppressed, with coefficients
\begin{align}
\text{family 1:}~~~\kappa_4, \kappa_5 \sim&   \dfrac{m_{3/2} }{M_{pl}^2} \epsilon^{10+ p} \simeq \dfrac{4 \times 10^{-20}}{M_{pl}} \Big(\dfrac{m_{3/2}}{10^3~\text{TeV}}\Big) \epsilon^{p} ~~\lesssim~~ \dfrac{10^{-8}}{M_{pl}} \\
\text{family 3:}~~~\kappa_4, \kappa_5 \sim&   \dfrac{m_{3/2} }{M_{pl}^2} \epsilon^{6+ p} \simeq \dfrac{3 \times 10^{-17}}{M_{pl}} \Big(\dfrac{m_{3/2}}{10^3~\text{TeV}}\Big) \epsilon^{p} ~~\lesssim~~ \dfrac{10^{-8}}{M_{pl}}~.
\end{align}
Flavor structure provides additional strong suppression, which depends on the considered family to which the particles belong and we have specified the minimum (family 3) and the maximum (family 1) possibilities. 

When coupled together,
the $R$-parity violating dimension-4 terms $\overline{U} \overline{D} \overline{D}$
 and $L Q \overline{D}$ can again lead to $p \rightarrow \overline{\nu}K^+$.
Thus, there is a constraint 
on the VEV the $\Phi$ field coming from the $\overline{U} \overline{D} \overline{D} \Phi$ and $L Q \overline{D} \Phi$ terms.
\begin{align} \label{eq:phivev}
\text{family 1:}~~~  \dfrac{\langle  \Phi \rangle}{M} \epsilon^{5 + 2p} \simeq 3 \times 10^{-4} \dfrac{\langle  \Phi \rangle}{M} \epsilon^{2p} ~ &\lesssim~10^{-13} \\
\text{family 3:}~~~ \dfrac{\langle  \Phi \rangle}{M} \epsilon^{2 + 2p} \simeq 4 \times 10^{-2} \dfrac{\langle  \Phi \rangle}{M} \epsilon^{2p} ~ &\lesssim~10^{-13}~.
\end{align}
We can neglect these conditions for the $\Phi$ field,
since the minimum of the potential \eqref{eq:potential_model}
is at $\langle \phi \rangle = 0$ if the $\phi^6$ term was absent.
Higher dimensional terms in the superpotential 
polynomial $F(x)$ can in principle modify this,
but they are highly suppressed.
Similarly, there is a constraint on the VEV of $\overline{\nu}$ from the $\overline{U} \overline{D} \overline{D} \overline{\nu}$ and $L Q \overline{D} \overline{\nu}$ terms, which appear non-pertubatively and thus carry extra suppression. These are
\begin{align} \label{eq:nuvev}
\text{family 1:}&~~~\Big(\dfrac{m_{3/2}}{M_{pl}}\Big) \dfrac{\langle \overline{\nu} \rangle}{M_{pl}}\epsilon^{5 + 2p}   \simeq  10^{-16} \Big(\dfrac{m_{3/2}}{10^3~\text{TeV}}\Big) \dfrac{\langle \overline{\nu} \rangle}{M_{pl}} \epsilon^{2p} ~ &\lesssim~10^{-13}~~ \\
\text{family 3:}&~~~\Big(\dfrac{m_{3/2}}{M_{pl}}\Big) \dfrac{\langle \overline{\nu} \rangle}{M_{pl}}\epsilon^{2 + 2p} \simeq  2 \times 10^{-14} \Big(\dfrac{m_{3/2}}{10^3~\text{TeV}}\Big) \dfrac{\langle \overline{\nu} \rangle}{M_{pl}} \epsilon^{2p} ~ &\lesssim~10^{-13}~.
\end{align} 
The condition from Eq.~\eqref{eq:nuvev}  on $\langle \overline{\nu} \rangle$ sets the scale for $(B - L)$ symmetry breaking  
\begin{equation}
M_{(B - L)} \lesssim  3 \, \epsilon^{-2p}~\Big(\dfrac{10^3~\text{TeV}}{m_{3/2}}\Big)  M_{pl}~,
\end{equation}
where we have assumed third family contribution, which is most restrictive since it has the smallest flavor suppression and thus leads to a lower $(B-L)$ scale. A possible fine-tuning of the $\overline{\nu}$ VEV can be easily avoided by
imposing that $(B - L)$ symmetry is unbroken to some lower energy scale
or alternatively that one or both of the contributing $\slashed{B}$ or $\slashed{L}$ operators
is suppressed or absent.

It is also possible that $\Delta B = 2$ neutron-anti-neutron oscillations \cite{Mohapatra:2009wp} can appear within the model with different mass-scale suppression than expected. Namely, upon integrating out the $\Phi$-field from the effective superpotential after SUSY/$R$ breaking
\begin{equation}
\mathscr{W}_{eff} \supset \dfrac{\epsilon^{5+2p}}{M}\overline{U}\overline{D}\overline{D}\Phi + M_{\Phi} \Phi \overline{\Phi} 
\end{equation}
one obtains the $(n - \overline{n})$ mediating operator
\begin{equation} \mathbf{O}_{n - \overline{n}} = \dfrac{\epsilon^{10+4p}}{M^2 M_{\Phi}} \Big( \overline{U} \overline{D} \overline{D}\Big)^2~,
\end{equation}
where in the above we have taken flavor structure corresponding to first generation that describes neutrons, which leads to the resulting operator being strongly suppressed. Similarly, after $\overline{\nu}$ acquires a VEV the $\overline{U} \overline{D} \overline{D} \overline{\nu}$ operator can lead to $(\overline{U} \overline{D} \overline{D})^2$, which just highlights the usual statement that $(B-L)$ scale can be related to $n - \overline{n}$ oscillations. However, since these terms appear non-perturbatively they are highly suppressed.
The free $n - \overline{n}$ lifetime is
\begin{equation}
\tau_{n-\overline{n}}^{\rm free} \sim \dfrac{1}{\delta m}~~~,~~~\delta m \simeq c \Big(\dfrac{\Lambda_{QCD}^4}{M^2 M_{\Phi}} \Big)\epsilon^{10+4p}~,
\end{equation}
where the right-hand side of the equation follows from dimensional analysis of the $\mathbf{O}_{n - \overline{n}}$ operator. We assume $\Lambda_{QCD} = 250$ MeV and $c \sim 1$. In the bound nuclei, the lifetimes differ. The free $n - \overline{n}$ lifetime can be converted to the bound $n - \overline{n}$ lifetime via
\begin{equation}
\tau_{n-\overline{n}}^{\rm bound} = R (\tau_{n-\overline{n}}^{\rm free})^2~,
\end{equation}
with $R \simeq 5 \times 10^{22}$ s$^{-1}$ being the nuclear suppression factor.
Hence, the current limits on bound $n - \overline{n}$ lifetime of $\tau_{n - \overline{n}}^{\rm bound} \gtrsim 2 \times 10^{32}$ years \cite{Abe:2011ky} imply 
\begin{equation} \label{eq:nnbarlife}
M_{\Phi} \gtrsim 10^{-1} \epsilon^{-4p} \Big(\dfrac{M_{pl}}{M}\Big)^2~\text{GeV}~.
\end{equation}
The constraint of Eq.~\eqref{eq:nnbarlife} is automatically satisfied in our model when cosmological bounds on reheating are taken into account through Eq.~\eqref{eq:mphi_reheat}, even for the most stringent case of $p = 2$.

\section{Baryon asymmetry and dark matter abundance from moduli decay}
\label{sec:cogenesis}

In our considerations we remain agnostic about the details of the inflationary phase. During inflation, moduli fields can develop
large values
and then dominate
the energy density of the universe as they coherently oscillate. 
If they are light and take too long to decay, they can on-set radiation dominated
universe at energies below the scale allowed by the BBN, resulting in
 a ``cosmological moduli problem'' \cite{Coughlan:1983ci,Banks:1993en}.
Making the moduli massive (``stabilizing moduli'') 
\cite{Giudice:1998xp,Randall:1998uk,Kachru:2003aw,Balasubramanian:2005zx,Conlon:2005ki}
helps to circumvent this issue. On other hand, massive moduli decaying to gravitinos can 
give rise to ``moduli-induced cosmological gravitino'' problem \cite{Ibe:2006am,Joichi:1994ce}. In our model both of these problems are avoided, as the moduli field $\Phi$ is massive and is forbidden to decay into gravitinos by the matter parity $\Z{2}$.  Here, unlike in the thermal cosmology, reheating radiation originates from decays of the $\Phi$ field and not the inflation. Since there are spontaneously broken discrete symmetries, resulting domain walls can be problematic. This issue is avoided, if the inflation scale is not extremely high and the domain walls have time to form and inflate away~\cite{Dine:2010eb}.
 
\subsection{Baryogenesis} 

The $\U{1}_A$ flavor structure implies that the $\overline{U}\overline{D}\overline{D}\Phi$ operator that drives baryogenesis also carries a flavor charge of
$5 + 2p$ (family 1 fields), $2 + 2p$ (family 2 fields) and $2p$ (family 3 fields), respectively. The contribution from the third family dominates. For $p = 0~(\tan \beta = 20)$ there is no suppression, while for $p = 1~(\tan \beta = 10)$ and $p = 2~(\tan \beta = 5)$ the operator is suppressed by $4 \times 10^{-2}$ and $2 \times 10^{-3}$, respectively. Thus, for $p \neq 0$ the resulting estimates for baryon asymmetry and DM abundance would need to be adjusted accordingly. For simplicity, we focus on the $p = 0$ case below.

Following \cite{Kitano:2008tk}, as in the Affleck-Dine mechanism \cite{Affleck:1984fy}, we define the $\phi$-number as
\begin{equation}
q_{\phi} = i (\dot{\phi}^{\ast} \phi - \phi^{\ast} \dot{\phi})~,
\end{equation}
which effectively denotes the difference in the number densities $n_{\phi}$ and $n_{\phi^{\ast}}$ of $\phi$ and $\phi^{\ast}$, respectively.
For a model of baryogenesis to be successful,
all three of the Sakharov's conditions \cite{Sakharov:1967dj} must be fulfilled: 
CP violation, baryon number violation ($\slashed{B}$) as well as out of
equilibrium interactions. Here, $q_{\phi}$-violating terms
in the potential generate $\slashed{B}$ and are also responsible for CP is violation. Since $q_{\phi}$-violation decouples, the condition of being out of equilibrium is also satisfied. 

The asymmetry in $\phi$ is generated by the suppressed $\phi^6$ K\"ahler potential term
\begin{equation}
K \supset X^{\dagger}X \Phi^6~,
\end{equation}
where $X$ is the spurion parametrizing SUSY/$R$ breaking. To preserve $q_{\phi}$ during coherent oscillations,  the $\phi$-violating terms should be sufficiently suppressed. Taken together with other terms, one obtains the following potential
\begin{equation} \label{eq:potential_model}
V = m_{\phi}^2 |\phi|^2 + m_{3/2}^2 M^2 F\Big(\dfrac{|\phi|^2}{M^2}\Big)
+ \Big[ \kappa \dfrac{m_{3/2}^2}{M^4} \phi^6 + h. c. \Big] + \dots
\end{equation}
where $F(x)$ is a general polynomial function,
$\kappa$ is a coupling that a priori is assumed to be of $\mathcal{O}(1)$ 
and dots represent higher order contributions. Here, SUSY/$R$  breaking 
has been taken to be of the gravitino mass size, with $m_X \sim m_{3/2}$.

The dynamics of $\phi$ field are described by the equation of motion
\begin{equation} \label{eq:phieom}
\ddot \phi + (3 H + \Gamma_{\phi}) \dot \phi + \dfrac{\partial V}{\partial \phi^{\ast}} = 0~,
\end{equation}
where $H$ is the Hubble parameter  and $\Gamma_{\phi}$ is
the $\phi$  decay rate. Treating $\phi$ as constant during $H \gg m_{\phi}$ 
allows to obtain the initial $q_{\phi}$ right after the end of inflation. Rewriting Eq.~\eqref{eq:phieom} as time evolution of the $\phi$-number gives
\begin{equation} \label{eq:phinumeom}
\dot q_{\phi} + 3 H q_{\phi} = - i \Big( \phi \dfrac{\partial V}{\partial \phi} 
- \phi^{\ast} \dfrac{\partial V}{\partial \phi^{\ast}} \Big) =
\dfrac{m_{3/2}^2}{M^4} \operatorname{Im}[\kappa \phi^6]~.
\end{equation}
Then, the initial condition for $q_{\phi}$ at $t_0 = 1 / m_{\phi}$ (i.e. when $H \sim m_{\phi}$)
is found to be
\begin{equation} \label{eq:phi_ini}
q_{\phi} (t_0) = |\kappa| \dfrac{m_{3/2}^2}{2 m_{\phi} M^4} \phi_{in}^6 \sim \dfrac{m_{3/2}^2 M^2}{m_{\phi}}~,
\end{equation}
where the initial $\phi_{in} = (m_{\phi} M^3)^{1/4}$ was taken as approximately $M$.
Assuming that $M$ is large, i.e. $\mathcal{O}(M_{GUT} - M_{pl})$, the $\phi$-asymmetry stays nearly constant
as $\phi$ starts to coherently oscillate. Namely,
the amplitude of $\phi$ in the right-hand side of Eq.~\eqref{eq:phinumeom} becomes small
relative to $M$. 
Hence,
the respective $\phi$-asymmetry is given by
\begin{equation}
\varepsilon \equiv \dfrac{q_{\phi}(t_0)}{n_{\phi} + n_{\phi^{\ast}}}
 \sim |\kappa| \Big( \dfrac{m_{3/2}}{m_{\phi}}\Big)^2~,
\end{equation}
where $n_{\phi}$ and similarly $n_{\phi^{\ast}}$ are given by 
$\rho_{\phi}/m_{\phi}$, with $\rho_{\phi} \simeq m_{\phi}^2 |\phi|^2 + |\dot \phi|^2$.

After oscillations $\phi$ decays (e.g. $\phi \rightarrow q q \tilde{q}$), reheating the universe and producing
superpartners. The $\phi$ decay rate is given by
\begin{equation}
\Gamma_{\phi} = \xi \dfrac{m_{\phi}^3}{M^2}~,
\end{equation}
where the numerical factor is $\xi = 27/(256 \pi^3) = 3 \times 10^{-3}$ \cite{Kitano:2008tk}, assuming all $\phi$ couplings to quarks as one.  
The reheat temperature $T_r$ is given by \cite{Thomas:1995ze}
\begin{equation}
T_r^2 \sim \sqrt{90 / g_{\ast} \pi^2} \Gamma_{\phi} M
\end{equation}  
where $g_{\ast}$ counts the effective number of relativistic
degrees of freedom and the whole numerical pre-factor is $\mathcal{O}(1)$  around the MeV scale, which corresponds to BBN. At  $H \simeq \Gamma_{\Phi}$, $T_r$ is 
found to be
\begin{equation} \label{eq:treh}
T_r \simeq 20~\text{GeV}~\Big(\dfrac{m_{\phi}}{5 \times 10^4~\text{TeV}}\Big)^{3/2}
 \Big(\dfrac{M_{pl}}{M}\Big)~.
\end{equation}
Requiring $T_r \gtrsim 4$ MeV to not spoil the BBN (e.g. \cite{Kawasaki:2000en})\footnote{During $\phi$ oscillation era, the $\phi$ energy density scales as matter $\rho_{m} \sim a^{-3}$,
in contrast to radiation energy density scaling as $\rho_{rad} \sim a^{-4}$. The scale factor $a$ can be taken as a measure of the 
universe's wavelength and  goes as $\sim 1/E \sim 1/T$. Hence, the higher the temperature $T_r$ for decoupling, the less BBN  dilution will be introduced by $\phi$.}, \eqref{eq:treh} constrains $m_{\phi}$ to be
\begin{equation} \label{eq:mphi_reheat}
m_{\phi} \gtrsim 150~\text{TeV} \Big(\dfrac{M}{M_{pl}}\Big)^{2/3}~.
\end{equation}

Assuming that $\phi$ dominates the energy density before its decay,
$T_r$ is related to $q_{\phi}$ via radiation density
\begin{equation}
\rho_{\phi} = m_{\phi} (n_{\phi} + n_{\phi^{\ast}}) \simeq \dfrac{\pi^2}{30} g_{\ast} T_r^4~.
\end{equation}
Since the baryon number density $n_b$ is set by $q_{\phi}$ via $n_b = q_{\phi}$,
the resulting baryon asymmetry is found to be (see Model A of \cite{Kitano:2008tk})
\begin{equation} \label{eq:baryogenesis}
\dfrac{n_b}{s} \simeq \varepsilon \dfrac{T_r}{m_{\phi}} \sim 
10^{-10}  |\kappa|  \Big(\dfrac{m_{3/2}}{10^3~\text{TeV}}\Big)^2
 \Big(\dfrac{5 \times 10^4~\text{TeV}}{m_{\phi}}\Big)^{3/2}
  \Big(\dfrac{M_{pl}}{M}\Big)~.
\end{equation}

\subsection{Dark matter}

With matter parity conserved, the lightest supersymmetric partner (LSP) is stable and thus constitutes the DM candidate, with a possible natural choice being the neutralino. Depending on the details of supersymmetry breaking, the sparticle spectrum will be altered and so will the identity of the neutralino that is the LSP.
Since we require a heavy gravitino of $m_{3/2} \gtrsim 50$ TeV for the scenario considered, loop corrections to soft masses (gauginos, etc.)
are relevant. This is a natural setting for anomaly-mediated
 supersymmetry breaking  \cite{Giudice:1998xp,Randall:1998uk} with
Wino DM \cite{Moroi:1999zb}. If the modulus and anomaly mediation contributions are competitive, Higgsino with mass below $\mathcal{O}$(TeV) becomes the DM candidate \cite{Allahverdi:2012wb}.
This is easily achieved in the context of ``mirage mediation'' \cite{Falkowski:2005ck,Choi:2007ka}, which could be further extended to ``deflected mirage mediation'' \cite{Everett:2008qy}, where the gaugino masses are given by
\begin{equation}
M_3 : M_2 : M_1 \sim (1-0.3 \alpha_m) g_3^2 : (1 + 0.1 \alpha_m) g_2^2 : (1 + 0.03 \alpha_m) g_1^2~.
\end{equation}
Here, $\alpha_m = m_{3/2}/ M_0 \log (M_{pl}/m_{3/2})$ parametrizes the relative strength of the anomaly and modulus-mediated contributions, with $M_0$ denoting the modulus-mediated contribution at the GUT scale, and $g_{1,2,3}$ are the gauge coupling constants. In the limit of $\alpha_m \rightarrow 0$, anomaly-mediated contribution vanishes and Bino becomes the DM candidate. The Higgsino mass parameter $\mu$ is driven by the gluino mass. Hence, increasing $\alpha_m$, which corresponds to increasing $m_{3/2}$ for a particular $M_0$, lowers the gaugino mass $M_3$ gluino component, while increasing the mass of the Bino $M_1$ and the Wino $M_2$ components, resulting in Higgsino DM. We treat $\alpha_m$ and $m_{3/2}$ as free parameters. 
 
Due to strong pair annihilation in the MSSM, Higgsino and Wino DM require non-thermal production,
which occurs due to the heavy $\phi$ decays. This sets
\begin{equation} \label{eq:dm_nontherm}
T_r \lesssim m_{\chi} / 20~,
\end{equation}
leading to (see Eq.~\eqref{eq:treh})
\begin{equation} \label{eq:chinonth}
m_{\chi} \gtrsim 40~\text{GeV}~\Big(\dfrac{m_{\phi}}{5 \times 10^4~\text{TeV}}\Big)^{3/2}
 \Big(\dfrac{M_{pl}}{M}\Big)~.
\end{equation}

From the Boltzmann equations, the relic density of $\chi$ is then approximately \cite{Fujii:2001xp}
\begin{equation}
\dfrac{n_{\chi}}{s} \sim (4 \langle \sigma v \rangle M_{pl} T_r)^{-1}~,
\end{equation}
where for particles carrying $\SU{2}$ quantum numbers, such as Wino or Higgsino, the thermally-averaged annihilation cross-section is $\langle \sigma v \rangle \sim 10^{-3}/m_{\chi}^2$. The resulting relic abundance \cite{Kitano:2008tk}
\begin{equation} \label{eq:dm_relic}
\Omega_{DM}  h_0^2 \simeq 0.1   \Big( \dfrac{m_{\chi}}{700~\text{GeV}}\Big)^3
 \Big(\dfrac{5 \times 10^4 ~\text{TeV}}{m_{\phi}}\Big)^{3/2}  \Big(\dfrac{M}{M_P}\Big)~,
\end{equation}
which is to be compared to the experimental value from \textit{Planck} \cite{Aghanim:2018eyx} 
of $\Omega_{DM} h_0^2 = 0.1198 \pm 0.0012$.

Hence, the DM and baryon abundances (Eq.~\eqref{eq:baryogenesis}) are related as
\begin{equation}\label{eq:dmtobaryon}
\dfrac{\Omega_{DM}}{\Omega_b} \sim 5~ |\kappa|^{-1}
  \Big(\dfrac{\text{1 GeV}}{m_{nuc}}\Big)
    \Big(\dfrac{m_{\chi}}{\text{700 GeV}}\Big)^3
  \Big(\dfrac{10^3~\text{TeV}}{m_{3/2}}\Big)^2   \Big(\dfrac{M}{M_P}\Big)^2~,
\end{equation} 
where $m_{nuc}$ is the nucleon mass. The $\Omega_{DM}/\Omega_b$ factor is required to be of $\mathcal{O}(1)$ if one is to address the observed baryon-DM abundance coincidence.

We first discuss Wino LSP, arising in the context of anomaly-mediated SUSY breaking. Here, the gaugino masses are proportional to
\begin{equation}
m_{G_i} \sim \dfrac{b_i g_i^2}{16 \pi^2} m_{3/2}~,
\end{equation}
where $G_i$ denote the SM gauge groups, $g_i$ are the gauge couplings and $b_i$ are the $\beta$-function coefficients. Hence,
\begin{equation}
m_{\chi} \simeq 2.7 \times 10^{-3} m_{3/2}~.  
\end{equation}
From \eqref{eq:dmtobaryon}, this results in
\begin{equation}\label{eq:winocoin}
\dfrac{\Omega_{DM}}{\Omega_b} \sim 70~ |\kappa|^{-1}
  \Big(\dfrac{\text{1 GeV}}{m_{nuc}}\Big)
    \Big(\dfrac{m_{\chi}}{\text{700 GeV}}\Big) \Big(\dfrac{M}{M_P}\Big)^2~.
\end{equation} 
It has been suggested (e.g. \cite{Fan:2013faa}) that
indirect detection already strongly constrains the non-thermal Wino. These results, however, strongly depend on the assumptions regarding the uncertain astrophysical $J$-factor associated with the DM halo shape in dwarf spheroidal galaxies \cite{Bhattacherjee:2014dya}. Assuming a more cored DM profile allows to alter the constraints by more than an order of magnitude in the $m_{\chi} \sim \mathcal{O}$(1) TeV parameter range. Indirect detection does provide a relatively robust constraint for $m_{\chi} < 300$, which in turn suggests a higher reheating temperature as well as a heavier modulus in the model. Future gamma-ray observations will allow to further probe the $m_{\chi} \sim \mathcal{O}$(1) TeV region. While indirect detection constraints are subject to large astrophysical uncertainties, disappearing track searches at colliders have proven to be a powerful probe for wino DM. Current Large Hadron Collider (LHC) searches already constrain wino with $m_{\chi} < 460$ GeV and upcoming high luminosity HL-LHC run studies will be sensitive to wino masses of up to 900 GeV \cite{Han:2018wus}. 

For non-thermal Higgsino, the mass region of $m_{\chi}$ is already also restricted \cite{Baer:2018rhs}. However, there exists an open parameter space region with $T_r \sim 20-50$ GeV and $m_{\chi} \sim 600-800$ GeV, which will be probed by the future Cherenkov Telescope Array (CTA) and 100 TeV collider experiments \cite{Aparicio:2016qqb}. For representative values of $T_r = 35$ GeV and $m_{\chi} = 700$ GeV, the  reheating (Eq.~\eqref{eq:mphi_reheat}, Eq.~\eqref{eq:treh}, Eq.~\eqref{eq:chinonth}), baryogenesis (Eq.~\eqref{eq:baryogenesis}), DM abundance (Eq.~\eqref{eq:dm_relic}) and DM-baryon coincidence (Eq.~\eqref{eq:dmtobaryon}) constraints are automatically satisfied for $m_{\phi} \simeq 5 \times 10^4$ TeV and $m_{3/2} \simeq 10^3$ TeV, assuming $M = M_{pl}$ as before.

We note that the mentioned DM constraints can be further alleviated by reducing the contribution of neutralinos to the overall DM abundance. This can be achieved, for example, within the frameworks where DM is a mix of axion(axino) and Higgsino \cite{Baer:2016hfa,Bae:2017hlp}.  As has been shown in \cite{Babu:2002ic}, this is realizable within a setting based on discrete gauge symmetries and can also lead to resolution of the strong CP problem via the Peccei-Quinn axion \cite{Babu:2002ic}. A full implementation of above into our model, however, is beyond the scope of the present work.~Other DM candidates, such as primordial black holes, have also been suggested in SUSY-based setting with scalar fields whose coherent-oscillations lead to an onset of a matter-dominated phase in the early universe~\cite{Cotner:2016cvr}.

\section{Conclusion}
\label{sec:conclusion}
 
The origin of the cosmological matter-antimatter asymmetry, the gauge hierarchy problem, the identity of the dark matter as well as the pattern of the fermion mass hierarchy and mixing angles remain among the greatest challenges for the Standard Model and Cosmology.
While minimal supersymmetric extension of the Standard Model remains the most promising solution for the gauge hierarchy problem, it has difficulty confronting the longevity of the nucleon lifetime and the smallness of the $\mu$ term. Further, the standard cosmological scenarios with thermal production of baryons and dark matter often face severe inconsistencies. 

In this work, we consider non-thermal moduli cosmology, which provides a consistent framework for generation of radiation, baryons, and dark matter, while maintaining successful BBN and avoiding the cosmological moduli problem. We construct an $\SU{5}$-compatible model based on discrete $\Z{12}^{R}$ $R$-symmetry combined with an anomalous $\U{1}_{A}$ flavor symmetry, in which the aforementioned problems are resolved simultaneously. Specifically, the $\Z{12}^{R}$ symmetry plays an important role in solving the $\mu$ problem, while the $\U{1}_{A}$ symmetry addresses the flavor puzzle. The interplay between $\Z{12}^{R}$ and $\U{1}_{A}$ symmetries leads to the absence or high suppression of proton decay operators as well as the prediction of Dirac neutrinos. The requirement of mixed $\U{1}_{A}$-gravitational anomaly cancellation automatically give rise to the correct size of the Dirac neutrino masses. The decay of the moduli $\phi$ field non-thermally produces baryons as well as dark matter, identified as the neutralino LSP. 
Consequently, the coincidence between the baryon number and dark matter abundance, $\Omega_{B} \sim \Omega_{DM}$, can be explained.  

\section*{Acknowledgments}

We thank Michael Ratz for participation during the initial stages of the project as well as Shigeki Matsumoto for discussion. The work of M.C.C. was supported by, in part, by the National Science Foundation under Grant No. PHY-1620638. The work of V.T. was
supported by the U.S. Department of Energy Grant
No. DE-SC0009937.
 
\bibliography{baryon}
\addcontentsline{toc}{section}{Bibliography}
\bibliographystyle{JHEP}

\end{document}